**Evidence for Clean d-wave Superconductivity in Samarium Nickelates**

Qingming Huang†[1], Xiaofang Fu†[2,3], Junlong Wu†[2,3], Laifeng Li*[4], Liang Qiao*[2,3] and Ye Yang*[1]

1. The State Key Laboratory of Physical Chemistry of Solid Surfaces, College of Chemistry & Chemical Engineering, Xiamen University, Xiamen 361005, China
2. School of Physics, University of Electronic Science and Technology of China, Chengdu, 611731, China
3. State Key Laboratory of Electronic Thin Films and Integrated Devices, University of Electronic Science and Technology of China, Chengdu 611731, China
4. State Key Laboratory of Cryogenic Science and Technology, Technical Institute of Physics and Chemistry, Chinese Academy of Sciences, Beijing 100190, China

† These authors contributed equally.

* lfli@mail.ipc.ac.cn; liang.qiao@uestc.edu.cn; ye.yang@xmu.edu.cn

**Abstract:** The discovery of superconducting nickelates provides a unique opportunity to explore the pairing mechanism of high-temperature superconductivity. Here, we use ultrafast terahertz spectroscopy to probe the temperature-dependent superfluid density in an infinite-layer samarium nickelate film with a $T_c$ of 20 K. The superfluid density decreases linearly with rising temperature, consistent with clean-limit d-wave pairing. From this linear relation, we extract a superconducting gap $\Delta(0) = 2.5 \pm 0.1$ meV and a gap-to-$T_c$ ratio, $2\Delta(0)\ /k_B T_c \approx 3$, suggesting that this sample lies in the weak-coupling regime. Furthermore, the ratio of the mean free path to the coherence length, $l/\xi$, is determined to be ~1.5, confirming clean-limit behavior. These findings establish strong parallels between the pairing mechanisms in nickelate and cuprate superconductors.

The discovery of superconductivity in infinite-layer nickelates has spurred intense research into their similarities with and distinctions from high-temperature cuprate superconductors.[1] One of the most important and enduring problems in these superconductors concerns the pairing mechanism, which is the key to understanding the nature and origin of the high-temperature superconductivity. Although nickelates share similar crystal and electronic structures with cuprates, critical questions regarding the electron pairing symmetry remain unsolved. Many theoretical studies on pairing instability predict a dominant $d_{x^2-y^2}$ pairing symmetry in the nickelate superconductors, analogous to cuprates, while others propose multiorbital nature.[2-10] Meanwhile, the experimental results provide a complex picture for the pairing symmetry. Scanning tunneling spectroscopy has revealed distinct gap structures consistent with both s-wave and d-wave pairing.[11] Temperature-dependent measurement of London penetration depth in different rare-earth nickelate films indicates both nodal d-wave and nodeless multigap pairing.[12,13] A recent terahertz (THz) spectroscopic study on a Sr-doped neodymium nickelate film suggests d-wave superconductivity in the dirty limit (i.e., mean free path $l$ < coherence length $\xi$).[14] Since the pairing feature in these measurements could be obscured in the presence of significant impurity scattering in the dirty limit, mimicking an alternative pairing mechanism,[12,13,15-18] establishing superconductivity in the clean limit (i.e., mean free path $l$ > coherence length $\xi$) is essential for disentangling intrinsic pairing properties from extrinsic influences. While clean-limit d-wave behavior is well established in cuprate superconductors,[16-23] experimental evidence for such a regime in nickelates remains lacking. Demonstrating clean-limit superconductivity in nickelates is thus a critical step toward clarifying their pairing symmetry and assessing the extent to which their superconducting mechanism parallels that of cuprates, with implications for a unified understanding of unconventional superconductivity in these systems.

The intriguing yet inconsistent results mentioned above underscore the need to clarify the pairing symmetry across the nickelate family. The recent achievement of significantly enhanced superconductivity in samarium nickelate films[24,25] renders them a particularly compelling system for such a study. However, due to current synthetic limitations, surface-sensitive techniques like angle-resolved photoemission spectroscopy face challenges in probing the superconducting state of the infinite-layer nickelates.[11,26,27] In contrast, bulk-sensitive techniques such as THz spectroscopy circumvent these issues and offer robust, precise methods to characterize the superfluid response across a wide temperature range.[12-14]

Here, we employ time-resolved optical pump-THz probe spectroscopy to investigate the photoinduced conductivity and photogenerated quasiparticle dynamics in a superconducting samarium nickelate film. The transient complex conductivity uncovers a photoinduced spectral weight transfer from superfluid to Drude response after optical excitation, a hallmark of Cooper pair breaking and quasiparticle formation. The recovery kinetics of the THz response are governed by bimolecular recombination of the photogenerated quasiparticles. Under weak excitation condition, the photo-destroyed superfluid density, which is proportional to the equilibrium suerpfluid density, exhibits a robust linear temperature dependence over a wide temperature range, consistent with clean-limit d-wave superconductivity. The ratio of $l/\xi$ is independently estimated to be around 1.5, which also satisfies the clean-limit condition. Our results demonstrate that nickelates can realize a clean superconducting regime, establishing a closer parallel to cuprates.

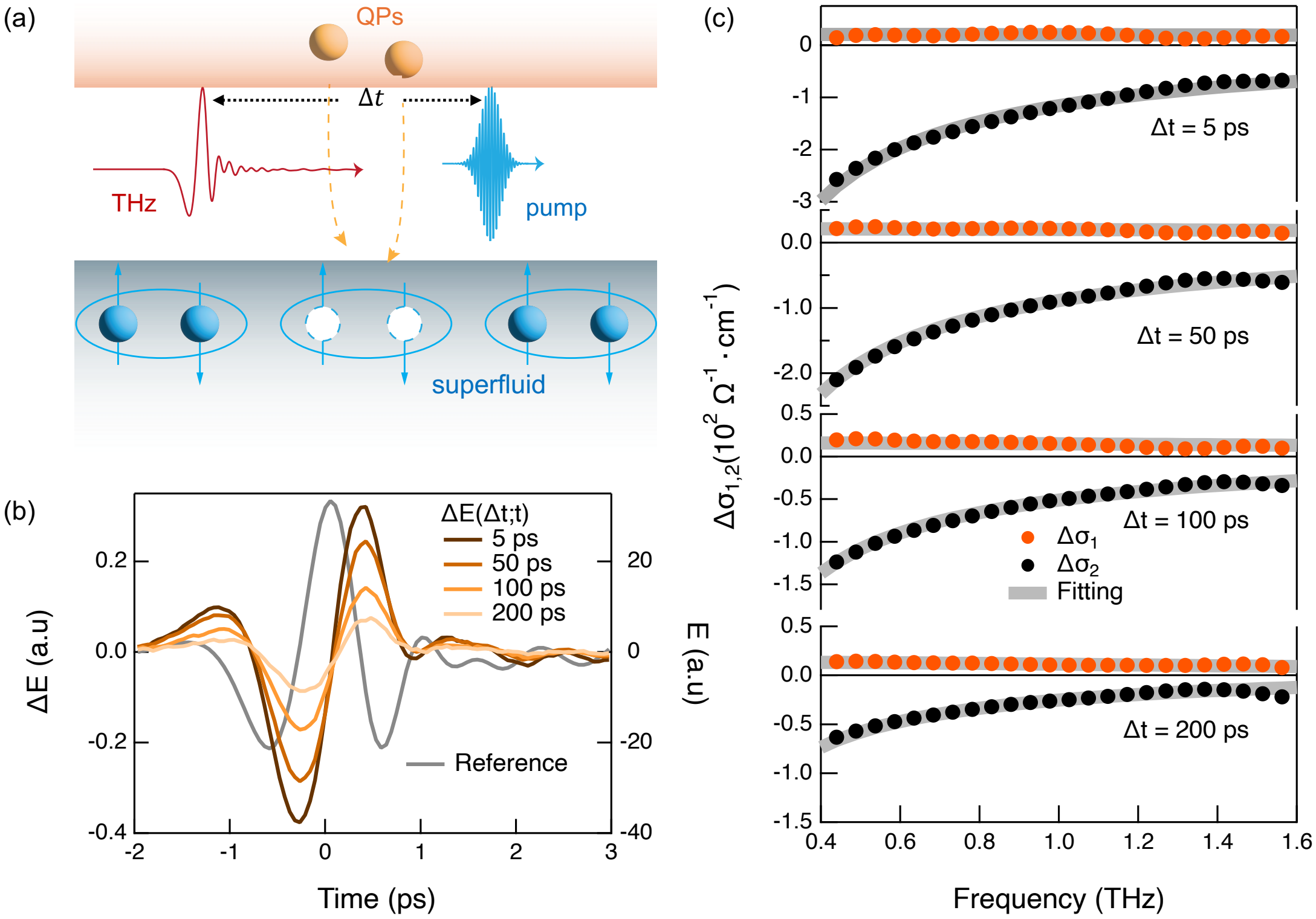

**Figure 1. Photoinduced THz conductivity change of nickelate SC films at 3 K.** (a) Schematic illustration optical excitation of condensed Cooper pairs. (b) Comparison between the reference THz waveform, $E(t)$, and signal waveforms, $\Delta E(t;\Delta t)$,recorded at representative pump-THz delays ($\Delta t$). (c) Real ($\Delta\sigma_1$, orange dots) and imaginary ($\Delta\sigma_2$, black dots) parts of the photoinduced change of the complex conductivity at indicated $\Delta t$. The gray dash curves represent the fitting curves based on the two-fluid model.

The $Sm_{0.75}Ca_{0.05}Eu_{0.2}NiO_2$ thin film with a thickness of 10 nm was epitaxially grown on a (La,Sr)(Al,Ta)$O_3$ (LSAT) substrate via pulsed laser deposition.[28] The electrical resistance of this nickelate film drops to zero at 20 K (Fig. S1), which is regarded as $T_c$. In the optical pump-THz probe measurements, a pump pulse ($h\nu$=1.55 eV) drives the superconducting nickelate film out of equilibrium, and a subsequent THz probe pulse measures the recovery process (Fig. 1a). The time delay ($\Delta t$) between pump and probe pulses was controlled by an optical delay line. Additional experimental details are provided in Supplementary Material. The transmitted THz electric field in the time domain, $E(t)$, where $t$ is the time delay between the THz generation and detection pulses, is recorded as a reference. As shown in Fig. 1b, the photoinduced change in this field, $\Delta E(t;\Delta t)$, show a significant temporal shift compared to $E(t)$. As $\Delta t$ increases, the amplitude of $\Delta E(t;\Delta t)$ decays progressively, whereas the shape of $\Delta E(t;\Delta t)$ remains unchanged, which indicates a clean recovery process without the formation of other transient species.

The frequency-dependent transient complex conductivity ($\Delta\tilde{\sigma}$) is extracted from the ratio of $\Delta\tilde{E}(\omega;\Delta t)/\tilde{E}(\omega)$, where $\Delta\tilde{E}(\omega;\Delta t)$ and $\tilde{E}(\omega)$ are the Fourier transforms of $\Delta E(t;\Delta t)$ and $E(t)$, respectively. At $T$ = 3 K, $\Delta\tilde{\sigma}$ displays a positive, Drude-like real part ($\Delta\sigma_1$) and a negative imaginary part ($\Delta\sigma_2$) that follows a $1/\omega$ dependence. This spectral signature indicates a photoinduced spectral weight transfer from the superfluid response to the Drude response, a hallmark of Cooper pair breaking and concurrent formation of quasiparticles, as previously observed in both nickelate and cuprate superconductors.[14,29-32] The $\Delta\tilde{\sigma}$ spectra for different $\Delta t$ are well described by a phenomenological two-fluid model (grey curves, Fig. 1c),[14,29,30,33] expressed as

$$\Delta\tilde{\sigma}(\omega) = \Delta\sigma_1 + i\Delta\sigma_2 = i\frac{\Delta\rho_s}{\omega} + \frac{\Delta\rho_n}{{\tau_n}^{-1} - i\omega} \qquad (1)$$

where $\Delta\rho_s$ and $\Delta\rho_n$ are the effective densities of broken Cooper pairs and the resulting quasiparticles, respectively, and ${\tau_n}^{-1}$ is the scattering rate of those photogenerated quasiparticles. The first term on the right side corresponds to the inductive ($1/\omega$) component associated with the photo-depleted superfluid density. In accordance with the spectral weight conservation, $\Delta\rho_n = -\Delta\rho_s$. As $\Delta t$ increases, both $\Delta\sigma_1$ and $\Delta\sigma_2$ exhibit a concurrent decay, reflecting the recovery of the superfluid density via the re-pairing of photogenerated quasiparticles. Throughout this study, the pump fluence was maintained in the low-excitation regime, where $\Delta\tilde{\sigma}$ scales linearly with the pump fluence (Fig. S2). This ensures that the system remains in the perturbative limit so that the number of broken Cooper pairs is proportional to the equilibrium superfluid density.[29,34] The low pump fluence also minimizes the influence of heating.

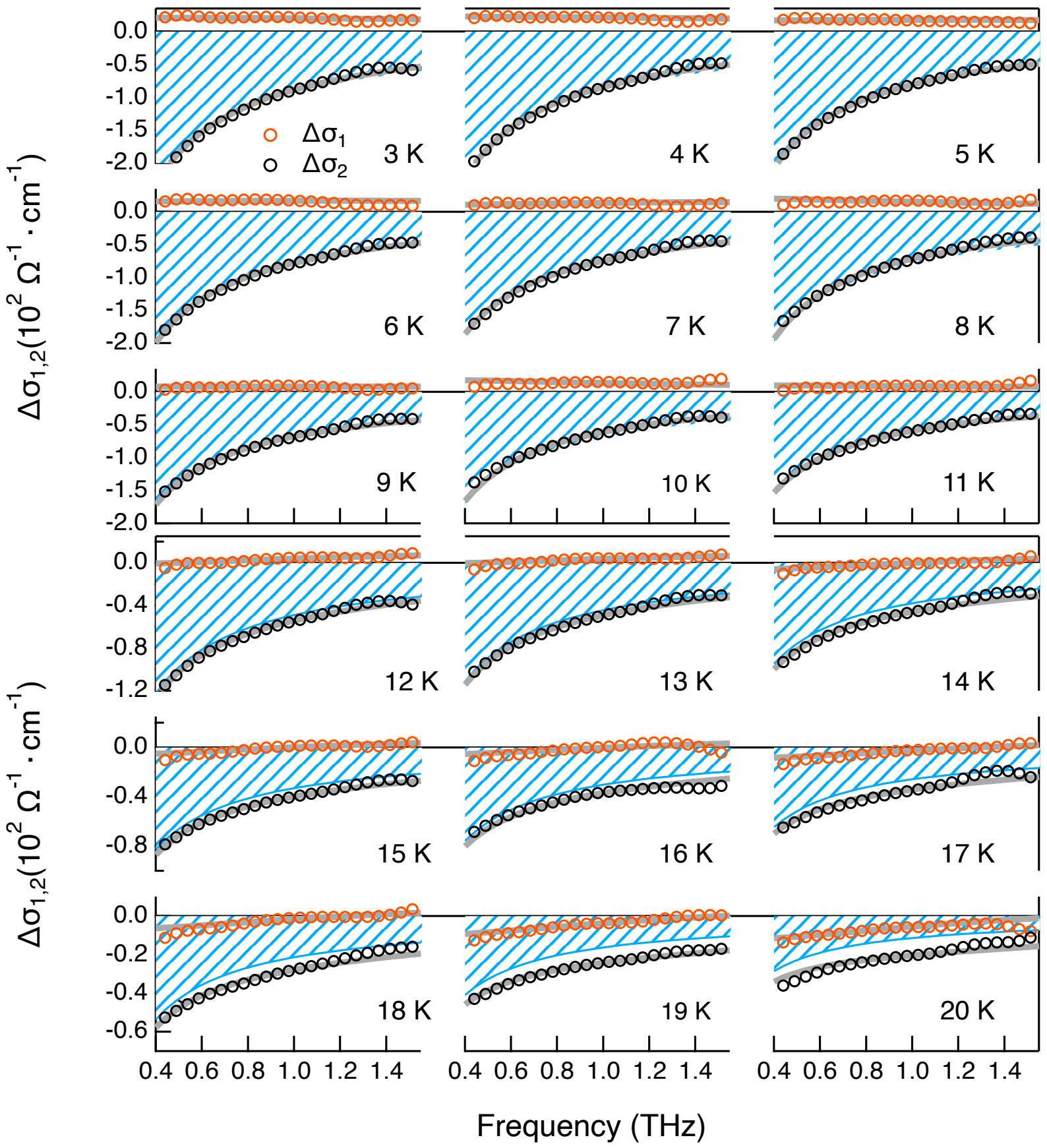

**Figure 2. Temperature dependence of the photoinduced THz conductivity change.** Real ($\Delta\sigma_1$, orange circles) and imaginary ($\Delta\sigma_2$, black circles) parts of the photoinduced change of the complex conductivities at various temperatures measured at a fixed delay of $\Delta t$ = 50 ps. The grey curves represent the fitting model. The shading represents contribution from $\Delta\rho_s$ (i.e., the first term on the right side of eq. 1).

To explore the temperature dependence of the superfluid density and pairing dynamics, we measured $\Delta\tilde{\sigma}$ at different temperatures. The data recorded at a fixed delay of 50 ps are displayed in Fig. 2. Under identical pumping conditions, the magnitude of both $\Delta\sigma_1$ and $\Delta\sigma_2$ decreases monotonically with increasing temperature, indicating a gradual reduction in $\Delta\rho_s$ and $\Delta\rho_n$. At temperatures well below $T_c$, the positive $\Delta\sigma_1$ and the negative $\Delta\sigma_2$ with its characteristic $1/\omega$ superfluid response are well described by the two fluid model. As the temperature approaches $T_c$, however, $\Delta\sigma_1$ partially crosses zero and eventually becomes fully negative, whereas $\Delta\sigma_2$ remains negative. This phenomenon has been attributed to the fluctuating superconductivity near $T_c$.[14]

The two-fluid model assumes that $\Delta\tilde{\sigma}$ arises solely from Cooper pair breaking, with its Drude component originating exclusively from the photogenerated quasiparticles, while thermal quasiparticles remain unperturbed.[29] However, previous work has shown that optical excitation may promote a small fraction of the thermal quasiparticles into a nonequilibrium "hot" state, reducing the thermal quasiparticle conductivity (i.e., normal conductivity) and thereby offsetting the positive contribution to $\Delta\sigma_1$.[14] This photoinduced perturbation of the thermal quasiparticle conductivity is particularly pronounced near $T_c$, which is probably linked to the fluctuations. An extended two-fluid model incorporating this perturbation (Eq. S2, SI) accurately reproduces $\Delta\tilde{\sigma}$ near $T_c$ (solid curves, Fig. 2).[14,35] At temperatures well below $T_c$, $\Delta\sigma_2$ contains negligible contribution from thermal quasiparticle excitation and is dominated by $\Delta\rho_s$ (shadings, Fig. 2a). In contrast, near $T_c$, the clear discrepancy between measured $\Delta\sigma_2$ (black circles, Fig. 2a) and the $\Delta\rho_s$ contribution (shadings, Fig. 2a) confirms appreciable thermal quasiparticle perturbation. $\Delta\tilde{\sigma}$ at other delays exhibits a similar temperature dependence (Figs. S3-S5), all well described by this extended two-fluid model. Notably, this extended two-fluid model remains phenomenological, and the microscopic mechanism linking the enhanced thermal quasiparticle perturbation near $T_c$ to superconducting fluctuations requires further investigation.

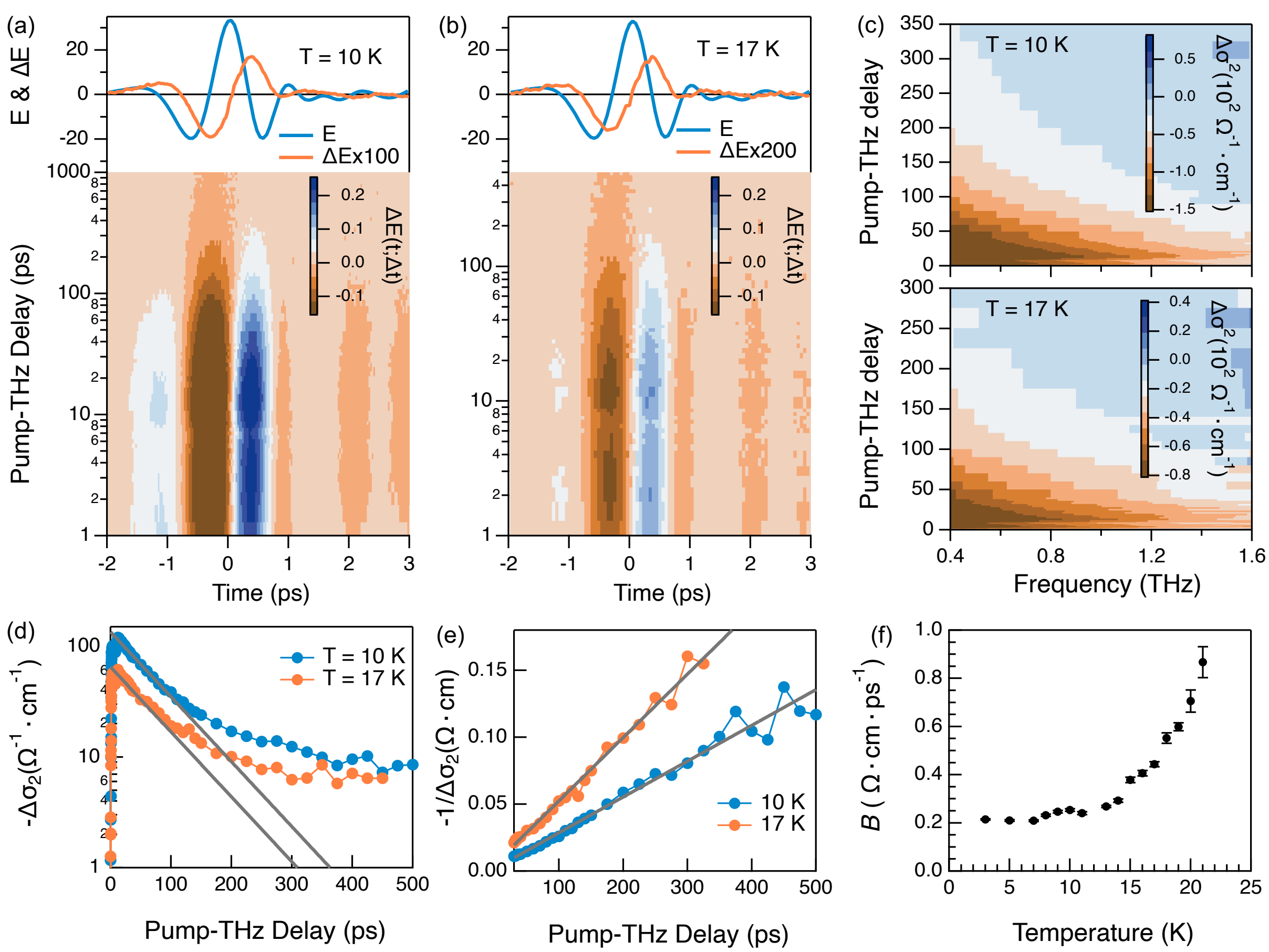

**Figure 3. Temperature dependence of the superfluid recovery dynamics.** (a)-(b) Time resolved THz waveforms measured at two different temperatures. The orange and cyan curves in top panels represent $\Delta E(t;\Delta t)$ and $E(t)$, respectively. (c) Time resolved $\Delta\sigma_2$ extracted from panel (a) and (b). (d) Kinetics of $\Delta\sigma_2$ at 1 THz at indicated temperatures. (e) The reciprocal of the $\Delta\sigma_2$ kinetics at indicated temperatures. (f) Temperature dependence of the bimolecular recombination rate ($B$) constant of the photogenerated quasiparticles.

The temporal evolution of THz waveforms at two representative temperatures shows a progressive decay in amplitude with negligible phase shift as $\Delta t$ increases (Fig. 3a-b), from which $\Delta\tilde{\sigma}$ at various $\Delta t$ can be extracted. The resulting temporal evolution of $\Delta\sigma_2$ (Fig. 3c) reveals a long-lived feature of the superfluid depletion that persists for several hundred picoseconds. This slow recovery of the superfluid density has previously been attributed to a bottleneck effect caused by the presence of superconducting gap.[34,36] According to the Rothwarf-Taylor (RT) model, superfluid recovery proceeds via the recombination of photogenerated quasiparticles into Cooper pairs,[29,37] ${dn^*}/{dt} \approx -B(n^*)^2 - 2Bn^*n_T$ , where $n^*$ and $n_T$ are the densities of the photogenerated and thermally excited quasiparticles, and $B$ is bimolecular recombination rate constant. This recombination can occur primarily through two pathways: bimolecular recombination between two photogenerated quasiparticles [i.e., $B(n^*)^2$ ], or between a photogenerated and a thermal quasiparticle (i.e., $2Bn^*n_T$).[29,37-40] The former gives rise to a bimolecular recovery, whereas the latter leads to a single-exponential recovery. If a quasi-equilibrium state is established between the quasiparticles and nonthermal phonons due to the strong electron-phonon coupling, the superfluid recovery dynamics become governed by the decay of this quasi-equilibrium state, typically limited by the nonthermal phonon lifetime, which potentially masks the quasiparticle recombination dynamics.[30,37,41]

Since $\Delta\sigma_2$ is dominated by $\Delta\rho_s$, its kinetics serve as a direct probe of the superfluid recovery dynamics. The kinetic traces at representative temperatures exhibit a strongly nonexponential decay, as evident from Fig. 3d, where y-axis is plotted on a logarithmic scale. The reciprocals of $\Delta\sigma_2$ kinetics increase linearly with $\Delta t$ (Fig. 3e), a hallmark of a bimolecular recombination process. This suggests that the recovery is governed by the mutual recombination between photogenerated quasiparticles [i.e., $B(n^*)^2$], a phenomenon previously observed in cuprate superconductors.[29,39] The $\Delta\sigma_2$ kinetics at all other measured temperatures also conform to this bimolecular pattern (Fig. S6). The bimolecular recombination rate constant $B$ extracted from the kinetics fitting is nearly temperature-independent for $T < 15$ K but rises steeply above 15 K. This temperature trend contrasts with observations in cuprates, where $B$ depends weakly on temperature,[39] likely because the participation of thermal quasiparticles at elevated temperatures causes the quasiparticle recombination in cuprates to significantly deviates from a pure bimolecular mechanism.[29] A systematic study will be needed to resolve this discrepancy.

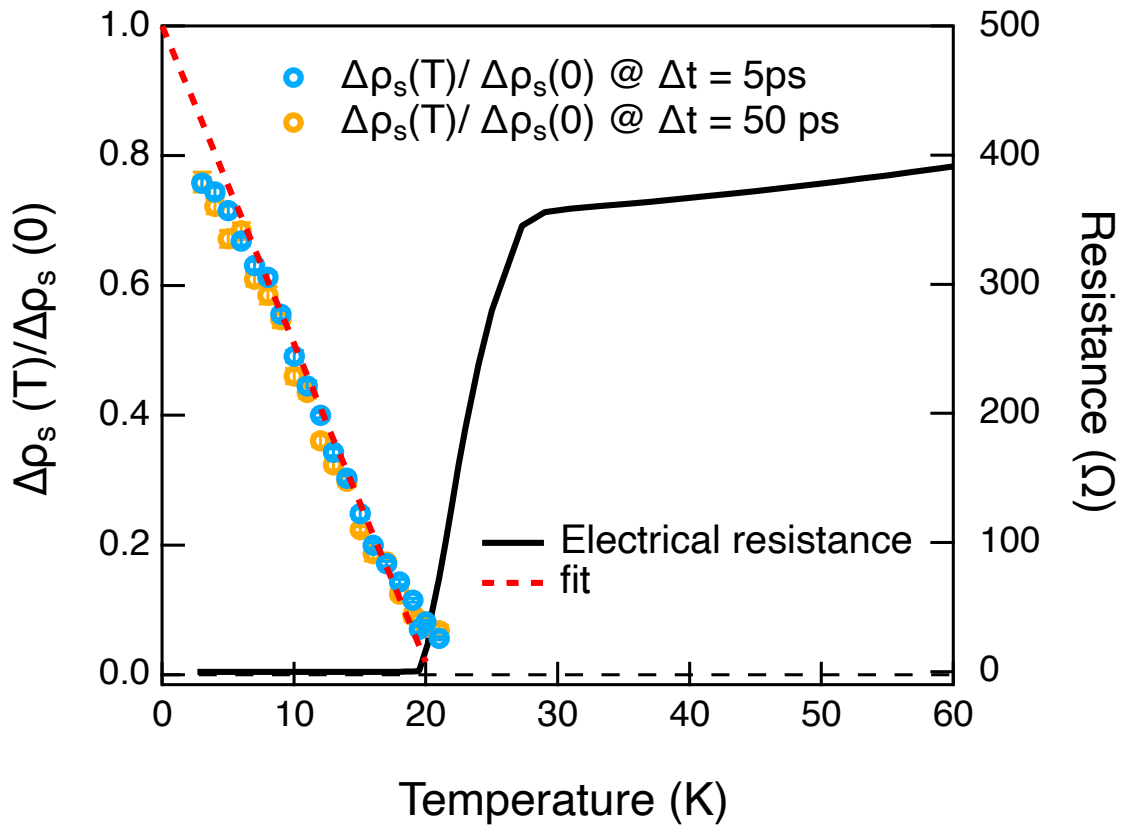

**Figure 4. Temperature-dependent superfluid density.** The cyan and orange circles represent the temperature-dependent $\Delta\rho_s(T)/\Delta\rho_s(0)$ at fixed pump-THz delay of 5 ps and 50 ps, respectively. The lines are the linear fits. The black curve is the temperature-dependent resistance.

The pairing symmetry in superconductors can be examined by measuring the temperature dependence of the normalized superfluid density, $\rho_s(T)/\rho_s(0)$ . A fully gapped s-wave superconductor exhibits an exponential temperature dependence of $\rho_s(T)/\rho_s(0)$.[13,16,42] In contrast, in nodal d-wave superconductors, the presence of line nodes leads to a linear temperature dependence of $\rho_s(T)/\rho_s(0)$ in the clean limit, which evolves into a power-law behavior (typically quadratic) in the presence of impurity scattering.[12,16] In other words, the linear temperature dependence of $\rho_s(T)/\rho_s(0)$ is a hallmark of clean-limit d-wave pairing, which is comprehensively validated in decades of cuprate superconductivity research.[17-22,43,44] Due to the low superfluid density, the equilibrium THz photoconductivity is dominated by the Drude response of the thermal quasiparticles (Fig. S7), and thus accurately extracting $\rho_s(T)$ based on the equilibrium measurement is very challenging. Nevertheless, under the perturbative excitation regime, the initially photo-destroyed superfluid density (the number of broken Cooper pairs) is proportional to the equilibrium superfluid density, i.e., $\Delta\rho_s(T) \propto \rho_s(T)$.[29,34] Thus, $\Delta\rho_s$ and $\rho_s$ should share the same temperature dependence. Given the relatively slow recombination below $T_c$ (Fig. 3d), $\Delta\rho_s$ recorded at short delays (e.g., 5 ps and 50 ps, see Fig. S8) can be considered as a close approximation to the initially photo-destroyed superfluid density.

As shown in Fig. 4, $\Delta\rho_s(T)/\Delta\rho_s(0)$ obtained at 5 ps and 50 ps are consistent with each other, and both display a characteristic linear temperature dependence. This linearity, extending from near $T_c$ (~18 K) down to 5 K, is in line with the clean-limit d-wave pairing, and it clearly does not reconcile with the quadratic temperature dependence for dirty-limit d-wave pairing or exponential temperature dependence for s-wave pairing. The linear temperature dependence of $\Delta\rho_s(T)/\Delta\rho_s(0)$ can be described by the following expression,[16,23]

$$\frac{\rho_s(T)}{\rho_s(0)} = \frac{\Delta\rho_s(T)}{\Delta\rho_s(0)} = \left[1 - \frac{2\ln 2}{\Delta(0)} k_B T\right] \qquad (2)$$

where $\Delta(0)$ is the superconducting gap magnitude at zero temperature. Below 5 K, the data begin to deviate from the linear trend, reminiscent of the crossover to a quadratic behavior in $\rho_s(T)$ that has been observed in cuprates at low temperature regime.[17,18,44] Therefore, the superconducting gap structure in this nickelate sample should be analogous to that in clean d-wave cuprate superconductors.

The linear fit of $\Delta\rho_s(T)/\Delta\rho_s(0)$ based on Eq. 2 (Fig. 4) yields $\Delta(0) = 2.5 \pm 0.1\ meV$, consistent with the value reported for Sr-doped neodymium nickelate film.[14] The gap-to-$T_c$ ratio, $2\Delta(0)\ /k_B T_c$, is determined to be around 3. This value is close to the weak-coupling d-wave Bardeen-Cooper-Schrieffer value in cuprate and nickelate superconductors,[14,45,46] and thus the measured nickelate film should also fall in the weak-coupling regime. The impurity influence on superconducting behaviors is often assessed by the ratio of $l/\xi = \pi\Delta(0)/\hbar\tau_0^{-1}$, where $\tau_0^{-1}$ is the scattering rate for the equilibrium quasiparticles. Fitting $\Delta\tilde{\sigma}(\omega)$ with the extended two-fluid model yields $\tau_0^{-1}$ = 8.3 THz, slightly higher than the value reported for the Sr-doped neodymium nickelate film.[14] Substituting $\tau_0^{-1}$ and $\Delta(0)$ into the expression above yields $l/\xi$~1.5, placing the system in proximity to the clean limit ($l > \xi$).[47] This result is compatible with the linear temperature dependence of $\Delta\rho_s$ over the wide temperature range.

Previous studies on lanthanum- and praseodymium-based nickelate superconductors with $T_c \leq 10$ K have reported a quadratic temperature dependence of $\rho_s(T)$, a characteristic signature of d-wave pairing in the dirty limit.[12,13] For neodymium-based nickelates, the literature remains divided. The scanning tunneling spectroscopy and penetration depth measurements have reported both nodal d-wave and nodeless s-wave behavior,[12,13,48] while pioneering terahertz work by Cheng et al. provided evidence for d-wave pairing in the dirty limit.[14] This inconsistence across the nickelate family suggest that variations in disorder and scattering may play a central role in shaping the observed superconducting response, masking the intrinsic nodal pairing symmetry in dirty-limit samples. In contrast, our result reveals the d-wave pairing in the clean limit in samarium-based nickelates. Demonstrating clean-limit superconductivity in nickelates strengthens their connection to cuprates, where d-wave pairing and nodal quasiparticles are well established. More generally, our results highlight the importance of quasiparticle coherence and scattering in determining superconducting properties in strongly correlated materials. The ability of time-resolved THz spectroscopy to selectively probe the superfluid response provides a broadly applicable approach for investigating superconductivity in systems where equilibrium measurements are hindered by low superfluid density or strong normal-state contributions.

In summary, we have investigated the pairing mechanism of infinite-layer samarium nickelate superconductors using ultrafast optical pump-THz probe spectroscopy. Temperature dependent measurements provide evidence for d-wave pairing symmetry in the clean limit, closely mirroring the behavior in cuprate superconductors.

**Acknowledgements**

This work was supported by the National Natural Science Foundation of China (52525208, 22473093), the National Key Research and Development Program of China (2023YFA1406301), the Key Research Program of the State Key Laboratory of Cryogenic Science and Technology (T-2025cryo-18), Science and Technology Department of Sichuan Province (2026NSFSCZY0033 and 2024ZYD0164).